\documentclass[superscriptaddress,twocolumn,aps,prl,floatfix,showpacs]{revtex4}
\usepackage{epsfig}
\usepackage{dcolumn}
\usepackage{bm}
\begin{document}
\title{An Intermediate Phase at the Metal-Insulator Boundary in a Magnetically Doped
Two-Dimensional Electron System}

\author{J.\ Jaroszy\'nski}\email{jaroszy@magnet.fsu.edu}
  \affiliation{National High
Magnetic Field Laboratory, Florida State University, Tallahassee,
FL 32310} 
\author{T. Andrearczyk}
 \affiliation{National High
Magnetic Field Laboratory, Florida State University, Tallahassee,
FL 32310} \affiliation{Institute of Physics, Polish Academy of
Sciences, 02--668 Warszawa, Poland}
\author{G. Karczewski}
\affiliation{Institute of Physics, Polish Academy of Sciences,
02--668 Warszawa, Poland}
\author{\\ J. Wr\'obel}
\affiliation{Institute of Physics, Polish Academy of Sciences,
02--668 Warszawa, Poland}
\author{T. Wojtowicz}
\affiliation{Institute of Physics, Polish Academy of Sciences,
02--668 Warszawa, Poland}
\author{Dragana Popovi\'c}
\affiliation{National High Magnetic Field Laboratory, Florida
State University, Tallahassee, FL 32310}
\author{T. Dietl}
\affiliation{Institute of Physics, Polish Academy of Sciences,
02--668 Warszawa, Poland} \affiliation{Institute of Theoretical
Physics, Warsaw University, 00-681 Warszawa,
Poland}\affiliation{ERATO Semiconductor Spintronics Project, Japan
Science and Technology Agency, 02--668 Warszawa, Poland}
\date{\today}

\begin{abstract}
A magnetotransport study in magnetically doped (Cd,Mn)Te 2D
quantum wells reveals an apparent metal-insulator transition as
well as an anomalous intermediate phase just on its metallic side.
This phase is characterized by colossal magnetoresistance-like
phenomena, which are assigned to the phase separation of the
electron fluid and the associated emergence of ferromagnetic
bubbles.
\end{abstract}


\maketitle

Despite intensive research efforts, the apparent
 metal-insulator transition (MIT)  in two-dimensional
  electron systems (2DES) \cite{abra01} remains one of the most
  challenging problems of condensed matter physics.
 Several recent theoretical studies \cite{chak99,dobr03,spiv04} suggest
 the existence of an intermediate phase
 near the 2D MIT,
  where the competition between distinct ground states  results
 in their nanoscale phase separation.  A large number of
possible configurations of these local regions often have
comparable energies, resulting in time-dependent phenomena such as
slow relaxation, aging, and other signatures of glassy dynamics.
Indeed, such manifestations of glassiness and evidence for an
intermediate metallic phase have been found recently in a 2DES in
Si metal-oxide-semiconductor field-effect transistors (MOSFETs)
\cite{bogd02,jaro04,jaro06}.
There is also growing evidence that phase separation is
responsible for several striking effects in bulk transition metal
oxides, such as manganites \cite{dago01}, cuprates \cite{schm00},
and similar complex magnetic materials~\cite{dago05}. In that
context,
magnetically doped 2DES (M2DES) in semiconductor heterostructures
constitute an ideal system for studying both magnetism and reduced
dimensionality effects near the MIT.

M2DES in (Cd,Mn)Te quantum wells (QWs) are particularly appealing,
since the 2D electron density $n_s$ can be changed externally by
an electric gate \textit{independent} of the density of magnetic
ions, so that interactions and the amount of disorder can be tuned
separately. These M2DES have been well characterized in the
studies of quantum Hall ferromagnetism~\cite{jaro02a} and quantum
Hall effect~\cite{Tera02}. Furthermore, (Cd,Mn)Te has a simple
crystal structure and, most importantly, extensive studies have
shown~\cite{Gaj94,shap84,DP-Dietl,jaro97acta} that the
distribution of Mn ions in (Cd,Mn)Te is perfectly random, implying
an \textit{absence} of chemical phase separation. Thus the
molecular-beam-epitaxy (MBE) grown (Cd,Mn)Te seems to be
chemically and structurally the cleanest system for studying a
possible electronic phase separation in a magnetic material.

Here we report a magnetotransport study of M2DES in (Cd,Mn)Te QWs,
which provides the first experimental evidence of an apparent MIT
in M2DES. At sufficiently high temperatures $T$, the MIT is
similar to that observed in low-mobility (high disorder) Si
MOSFETs~\cite{bogd02}. However, when $T$ is low enough, a
qualitatively new transport regime emerges just on the metallic
side of the MIT. In this intermediate phase, the resistivity
$\rho$ increases
dramatically by several orders of magnitude with decreasing $T$.
A magnetic field $B$ applied \textit{parallel} to the 2D plane
gives rise to
an enormous \textit{negative} magnetoresistance (MR), which drives
the system back to metallic conductivity. This is exactly the
opposite of the behavior observed in nonmagnetic 2DES near the
MIT~\cite{abra01}, but it resembles the phenomena found in
manganites and other colossal magnetoresistance (CMR)
materials~\cite{dago01}. We propose that they share essentially
the same origin: a nanoscale coexistence of competing phases
which, in our case, corresponds to the formation of ferromagnetic
(FM) metallic bubbles embedded within a carrier-poor, magnetically
disordered host.
This picture is further supported by the observed strongly
nonlinear current-voltage ($I$-$V$) characteristics in this novel
regime, where sufficiently high excitation voltages $V_{exc}$
destroy the heterogeneous state and, hence, drive the system back
to metallic conductivity.

Our MBE grown samples
contain a 10~nm wide Cd$_{1-x}$Mn$_x$Te QW, in which the 2DES is
confined by Cd$_{0.8}$Mg$_{0.2}$Te barriers~\cite{jaro02a}.
A 10~nm thick layer of the front barrier residing ~20~nm away from
the QW is doped with iodine donors up to $n_I\approx
10^{18}$~cm$^{-3}$. In Structures A and B, Mn contents $x$ are
0.015 and 0.005, the peak 4.2~K mobilities $3.0\mbox{ and
}6\times10^{4}$ cm$^2$/Vs,
$1.4\le n_s \le 4.2\mbox{ and }4.9 \le n_s \le 6.2\times 10^{11}$
cm$^{-2}$ (controlled by a metal front gate), respectively. The
experiments are performed in parallel $B$ up to 9~T, and down to
either 0.24~K in a pumped $^3$He system or 0.03~K in a
$^3$He/$^4$He dilution refrigerator.
$\rho(B,T)$ is measured employing a standard low-frequency lock-in
technique.

\begin{figure}[t]
\includegraphics[width=7.9cm]{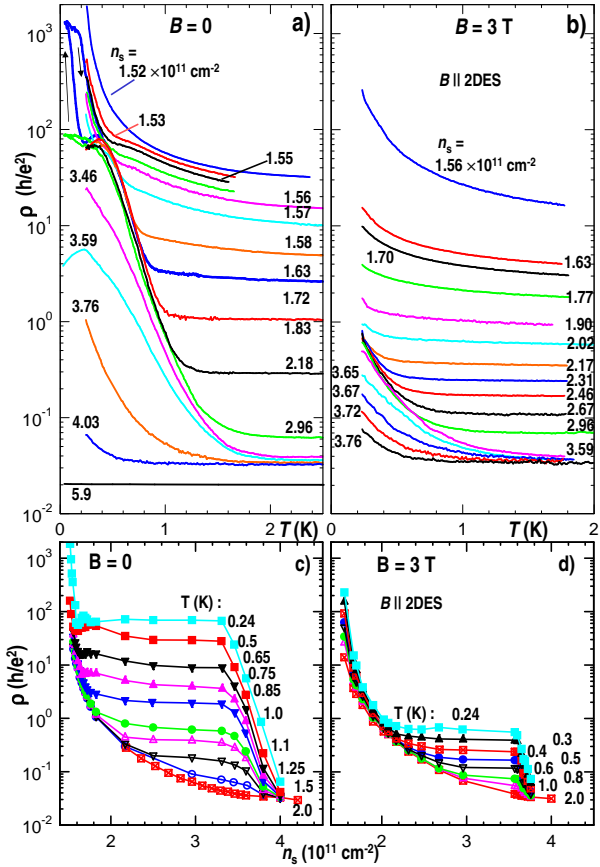}
\caption {(color online) Resistivity as a function of $T$ for
different electron densities at $B=0$ (a) and $B=3$~T (b). The
same data as a function of $n_s$ at selected $T$ (c, d). Except
for the lowest trace in (a) all the data are from Structure
A.}\label{fig:RvT}
\end{figure}

Figures~\ref{fig:RvT}(ab) show  $\rho(T)$ of Structure A at $B=0$
and $B=3$~T for different $n_s$.
The data were obtained with low
$V_{exc}\sim 10$~$\mu$V.  At elevated $T$ and high $n_s$,
$\rho(T)$ is weakly metallic, {\it i.e.} $d \rho/d T\gtrsim0$,
which is seen better in the scale of Fig.~\ref{fig:mit_iv}(b).
 At $n^*_s\approx
2.4\times10^{11}$ cm$^{-2}$, $d\rho/dT$ changes sign, which is
sometimes attributed to an apparent 2D MIT. In our case,
this occurs at a relatively low $\rho\sim 0.2$~$h/e^2$. The 2D
metallic behavior in a (Cd,Mn)Te QW is similar to that observed in
n-GaAs \cite{hane98} as well as in low-mobility Si
MOSFETs~\cite{bogd02}, where the $\rho(T)$ dependence in the
metallic phase is also weak. These are in contrast to a rather
strong $\rho(T)$  found in high-mobility Si MOSFETs~\cite{krav94}.
However, the critical density $n_c$ for the MIT obtained from the
extrapolation of the hopping activation energy, as well as from
the saturation of the $I$-$V$ characteristics \cite{shas01} is
substantially lower:
$n_c=1.7\times10^{11}$ cm$^{-2}<n_s^*$,
corresponding to $\rho_c\sim 2 h/e^2$, similar to other 2D
systems. At $B=3$~T, we find $n_c(B\mbox{=3 T})=2.1\times10^{11}$
cm$^{-2}$, indicating that, for high enough $T$, $B$ shifts the
system towards an insulating phase just as in standard
non-magnetic materials.

However, below some $T^*(n_s)$, a dramatic upturn of $\rho(T)$ by
almost three orders of magnitude is observed for moderate $n_s$ at
$B=0$ [Fig.~1(a)]. Here, $\rho(T)$ increases down to
$T_C\approx0.3$~K, goes through a maximum, and either continues to
grow or decreases as $T$ is lowered. Around $T_C$, a strong
resistance noise is observed (not shown), while below  $T\lesssim
T_C$, $\rho$ becomes hysteretic with respect to both $T$ and $B$.
Moreover, in the range of $T_C<T<T^*$, $\rho(T)$ for a wide range
of densities $1.6\lesssim n_s\lesssim 3.3\times10^{11}$ cm$^{-2}$
collapse onto almost the same curve. This striking behavior is
further highlighted in Fig.~\ref{fig:RvT}(c), which shows clearly
that $\rho$ does \textit{not} depend on $n_s$ at a fixed $T$.
Importantly, these densities $n_s\gtrsim n_c=1.7\times10^{11}$
cm$^{-2}$, i.e. they belong to the metallic side of the MIT. The
same dramatic upturn of $\rho$, as well as a lack of $\rho(n_s)$
dependence, are also clearly seen at $B=3$~T
[Figs.~\ref{fig:RvT}(bd)] at a slightly lower $T^*$ and a narrower
range of densities $2.1\lesssim n_s\lesssim 3.0\times10^{11}$
cm$^{-2}$. Again, these $n_s\gtrsim n_c(B=3\mbox{
T})=2.1\times10^{11}$ cm$^{-2}$.

\begin{figure}[tb]
\includegraphics[clip,width=7.9cm]{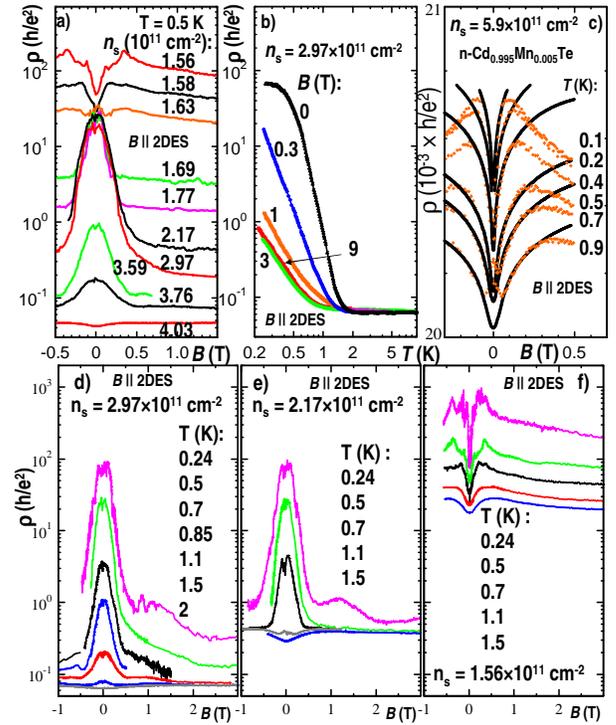}
\caption {(color online) (a) Resistivity $\rho(B)$   for different
$n_s$ at $T=0.5$~K. (b) $\rho(T)$ at different $B$. (c)  $\rho(B)$
at different $T$ measured (symbols) and calculated (lines) for
Structure B. (d), (e), and (f) $\rho(B)$ measured at different $T$
for $n_s=2.96,\;2.17\;\mbox{ and }1.56 \cdot10^{11}$~cm$^{-2}$,
respectively.}\label{fig:MR1}
\end{figure}

Figure 2 depicts the MR of our devices.  At high $T$, and at low
$T$ for either high or low $n_s$, a relatively weak positive MR
is visible. In the weakly localized regime, where $k_F\ell > 1$
($k_F$ -- Fermi wave vector, $\ell$ -- mean free path), this
positive MR is ubiquitous in diluted magnetic semiconductors
(DMS), such as (Cd,Mn)Te, in the paramagnetic phase
\cite{Sawi86,Andr02,smor97,Andr05}. It originates from the giant
spin-splitting $\Delta_s$ of the electron states, which
considerably affects quantum corrections to the conductivity
brought about by the effect of disorder modified electron-electron
interactions \cite{Alts85Fuku85Lee85}. The same mechanism is
believed to be responsible for positive MR in Si MOSFETs
\cite{dass05} and other nonmagnetic 2DES, in which $\Delta_s$ is
appropriately large. Since in DMS $\Delta_s$ is proportional to
the magnetization $M$ of the Mn spins, this positive MR scales
with $B$ and $T$ like the Brillouin function ${\textrm
B}_{5/2}[B/(T+T_{AF})]$~\cite{DP-Dietl}. Here $T_{AF}(x)>0$
describes a reduction of $M$ by intrinsic antiferromagnetic (AF)
couplings between the Mn ions, and its value is well known from
extensive magnetooptical and magnetic studies in (Cd,Mn)Te layers
\cite{Gaj94}. In particular, $T_{AF}=0.5$~K for $x =0.015$.
Figure~\ref{fig:MR1}(c) shows the results of the MR calculations
carried out according to this model
\cite{Alts85Fuku85Lee85,Sawi86,Andr02} and without any fitting
parameters, together with the experimental results for Structure B
in which $k_F\ell >> 1$, so that the theory should apply
quantitatively. Indeed, there is a good agreement between measured
and calculated positive MR at low $B$.

However, the overall MR is dominated by a strong negative
component. Importantly, this CMR shows up exclusively in the range
of $T$ and $n_s$, where the lowering of $T$ results in a strong
increase of $\rho(T)$. The effect is not observed on the
insulating side of the MIT, where localization is driven by
non-magnetic disorder and visible already at high $T$. A sizable
negative MR shows up only if magnetic effects lead to
$T$-dependent localization, so that $B$ can drive the system back
to the metallic phase. Remarkably, this negative MR does not scale
with $M(T,B);$ a relatively weak field is sufficient to destroy
the magnetism-related localization. As a result, a colossal
negative MR  is observed.

Similar effects, in particular, a dramatic upturn of $\rho(T)$ and
the associated negative MR, have been observed in bulk
DMS~\cite{Sawi86,Glod94,Leig98,Andr05}
below a characteristic temperature $T^*$ ($T^{*}\approx 1.5-2.0$~K
in n-Cd$_{1-x}$Mn$_x$Te~\cite{Leig98}).  They were
attributed~\cite{Sawi86,Leig98} to the formation of bound magnetic
polarons (BMP), even though the region where those effects are
observed spreads from the vicinity of the MIT deep into the
metallic phase, in remarkable analogy to the 2D system.  In
addition, in a modulation-doped n-(Cd,Mn)Te QW, Mn are neutral,
while the ionized donors are far from the conducting channel, so
that the BMP effects are expected to be virtually absent or, at
least, considerably less important than in 3D.
On the other hand, the tendency towards phase separation should be
more pronounced in 2D than in 3D~\cite{Imry75}.
We propose, therefore, that similarly to the case of
manganites~\cite{dago01}, the novel intermediate phase in M2DES
contains bubbles of different electronic phases,
which account for the CMR-like behavior.

In order to identify the nature of the phases in question, we note
that, according to the Zener theory of carrier-mediated FM, a
ferromagnetic transition is expected to occur at low $T$ in bulk
zinc-blende DMS~\cite{diet00}, such as (Cd,Mn)Te.
Recent Monte Carlo simulations~\cite{dp-alva02,yu05} indicate a
formation of isolated ferromagnetic bubbles and the CMR-like
behavior
at $T^*>>T_C$ ($T_C$ -- Curie temperature) even in the absence of
attractive impurity potentials, provided that AF couplings are
strong enough. At $B=0$,  the FM bubbles are oriented randomly,
which diminishes percolation and thus enhances resistance. Since
the magnetic field  aligns the bubbles, a strong negative MR
follows. These effects are expected to exist in both 3D and 2D
cases but should be stronger in 2D, in agreement with our results.
The lack of $\rho (n_{s})$ dependence in the intermediate phase
[Fig.~\ref{fig:RvT}(c)], where magnetic effects take control over
charge transport, is also consistent with the Zener theory of
ferromagnetism \cite{diet00,diet05}, which predicts that FM
ordering temperature depends merely on the carrier density of
states (DOS) and magnetic susceptibility of the Mn ions. However,
DOS does not depend on $n_s$ in 2D, resulting in this striking
behavior. At the lowest $n_s$, where carriers become localized,
the local FM order is destroyed, and the intrinsic AF interactions
between the Mn ions dominate \cite{diet05}.

\begin{figure}[tb]
\includegraphics[width=6.8cm]{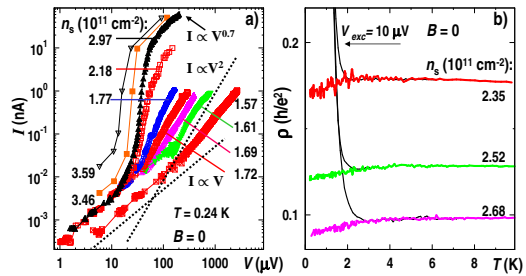}
\caption {(color online) (a)  $I$-$V$ characteristics  at
different $n_s$. (b) $\rho(T)$ for different $n_s$ at high (thick)
and low (thin lines) excitation voltages $V_{exc}=$ 500 and 10
$\mu$V, respectively.}\label{fig:mit_iv}
\end{figure}

In general, a heterogeneous state may be expected to exhibit
glassy behavior and nonlinearities~\cite{dago05}.  Indeed,
signs of glassiness, such as $\rho$ noise and hysteretic behavior
[Fig.~\ref{fig:RvT}(a)] are present in our system.  We have
established that the $I$-$V$ characteristics are strongly
nonlinear [Fig.~\ref{fig:mit_iv}(a)]. In particular,
while the Ohmic regime is observed for all values of $n_s$ for
$V_{exc}$
below $\approx10$~$\mu$V, a strong nonlinear behavior is clearly
seen for higher $V_{exc}$.  For low $n_s$,  the nonlinear
characteristics obeys $I\propto V^2$, but in the intermediate,
heterogeneous state,
a much steeper dependence is first observed, which is then
followed by $I\propto V^{0.7}$.  This striking behavior resembles
depinning of colloids \cite{dp-reic02}, Wigner glass to liquid
transitions modeled for disordered 2DES~\cite{reic04},
as well as Wigner crystal depinning in quantum Hall
systems~\cite{gold90}. In particular, the exponents 2 and 0.7 were
found \cite{dp-reic02} in the $I$-$V$ characteristics of colloidal
dynamics and correspond to plastic and elastic depinning,
respectively.  It is also possible that the observed nonlinearity
stems from the current-induced rotation of FM domains, though the
current density used here is 3 orders of magnitude lower than that
employed for domain rotation in (Ga,Mn)As \cite{DP-ohno02}.  A
sufficiently high
$V_{exc}$ destroys the heterogeneous state and drives the system
back to metallicity, as expected: for $V_{exc}
>100$~$\mu$V~$\gtrsim T^*$, the metallic behavior ($d \rho/d T>0$) is clearly seen down to the lowest $T$ for
$n_s\gtrsim2.2\times10^{11}$~cm$^{-2}$ [Fig.~\ref{fig:mit_iv}(b)].

\begin{figure}[tb]
\includegraphics[width=6.8cm]{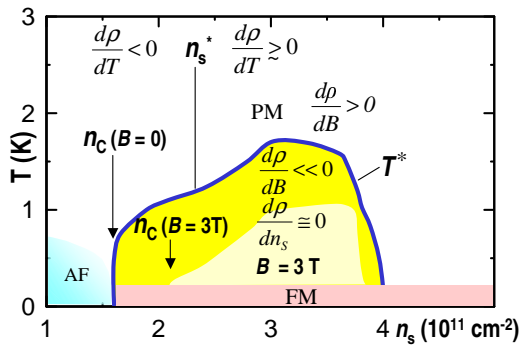}
\caption {(color online) Different transport regimes in the
$n_s$--$T$ plane determined from the  measurements in Structure A.
FM ordering is expected in the metallic phase at low enough $T$
based on the studies of 2D p-(Cd,Mn)Te QW~\cite{Bouk02,diet05} and
3D p-(Zn,Mn)Te~\cite{Ferr01,diet00}, although other states might
be possible.  For $n_s<n_c$, we expect AF correlations to
dominate.}\label{fig:mit}
\end{figure}

The different transport regimes observed in our M2DES are shown in
Fig.~\ref{fig:mit}.  We note that, even if the BMP model could
apply for $n_c<n_s<n_{s}^{*}$, one would expect a decrease of
$T^*$ with the increasing $n_s$, resulting from a dramatic
increase or a divergence of the localization length as
$n_s\rightarrow n_{s}^{*}$. This is exactly the opposite of what
is observed in the experiment.  On the other hand, it is expected
in general~\cite{dobr03} that a heterogenous state may exist in
the metallic phase just above the MIT: as the density of carriers
$n_s$ and their Fermi energy $E_F$ are reduced, any tendency
towards magnetic ordering will be revealed just prior to the MIT.
Since FM correlations are mediated by mobile carriers, the local
magnetic ordering and the heterogeneous phase will disappear as
the number of mobile carriers is further reduced to zero as
$n_s\rightarrow n_c$ from above, in agreement with our results.
In contrast, BMPs are most stable within the localized phase.  It
is interesting that the phase diagram (Fig.~\ref{fig:mit}) reveals
a striking qualitative similarity to that proposed for CMR
materials \cite{dago01}, cuprates \cite{dobr05} and, in general,
for model systems that form heterogeneous phases \cite{reic05}.

In summary, we have observed the emergence of an anomalous
intermediate phase exhibiting CMR-like phenomena in a M2DES just
on the metallic side of the MIT.  We attribute our findings to the
competition between AF exchange characterizing the insulating
phase and the FM correlations induced by itinerant electrons,
resulting in the formation of
FM metallic bubbles embedded within a carrier-poor, magnetically
disordered matrix.  Similarities to other systems, including
nonmagnetic 2DES
\cite{bogd02,jaro04,jaro06}, suggest a common origin of complex
behavior near the MIT in a variety of materials.

We thank P. Littlewood and V. Dobrosavljevi\'c for stimulating
discussions. This work was supported by INTAS Grant No.
03-51-5266, NSF Grants No. DMR-0071668 and 0403491, and NHMFL
through NSF Cooperative Agreement No. DMR-0084173.

\end{document}